\documentclass[10pt, conference]{IEEEtran}
\IEEEoverridecommandlockouts
\usepackage{cite}
\usepackage[hyphens]{url}
\usepackage{breakurl}
\usepackage[colorlinks,citecolor=black, filecolor=black, linkcolor=black,urlcolor=black]{hyperref}
\usepackage{flushend}
\usepackage{makecell}
\usepackage{subfigure}
\usepackage{amsmath,amssymb,amsfonts,bm}
\usepackage{algorithmic}
\usepackage {algorithm}
\usepackage{graphicx}
\usepackage{textcomp}
\usepackage{float}
\usepackage{multirow}
\usepackage{booktabs}
\usepackage[normalem]{ulem}
\useunder{\uline}{\ul}{}
\usepackage{xcolor}

\usepackage{geometry}

\def\BibTeX{{\rm B\kern-.05em{\sc i\kern-.025em b}\kern-.08em
    T\kern-.1667em\lower.7ex\hbox{E}\kern-.125emX}}

\begin{document}

\title{Alternate Learning based Sparse Semantic Communications for Visual Transmission\\
\thanks{This work was supported in part by the National Natural Science Foundation of China under Grants 62071425, in part by the Zhejiang Key Research and Development Plan under Grant 2022C01093, in part by Huawei Cooperation Project, and in part by the Zhejiang Provincial Natural Science Foundation of China under Grant LR23F010005.}
}

\author{Siyu Tong, Xiaoxue Yu, Rongpeng Li, Kun Lu, Zhifeng Zhao, and Honggang Zhang\\
\thanks{S. Tong, X. Yu, and R. Li are with College of Information Science and Electronic Engineering, Zhejiang University (email: \{tongsiyu, sdwhyxx, lirongpeng\}@zju.edu.cn).} 
\thanks{K. Lu was with Zhejiang University and is now with Huawei Technologies Co., Ltd. (email: lukun199@zju.edu.cn).}
\thanks{Z. Zhao and H. Zhang are with Zhejiang Lab as well as Zhejiang University (email: \{zhaozf, honggangzhang\}@zhejianglab.com).}
}
\maketitle

\begin{abstract}
Semantic communication (SemCom) demonstrates strong superiority over conventional bit-level accurate transmission, by only attempting to recover the essential semantic information of data. In this paper, in order to tackle the non-differentiability of channels, we propose an alternate learning based SemCom system for visual transmission, named SparseSBC. Specially, SparseSBC leverages two separate Deep Neural Network (DNN)-based models at the transmitter and receiver, respectively, and learns the encoding and decoding in an alternate manner, rather than the joint optimization in existing literature, so as to solving the non-differentiability in the channel. In particular, a ``self-critic" training scheme is leveraged for stable training. Moreover, the DNN-based transmitter generates a sparse set of bits in deduced ``semantic bases", by further incorporating a binary quantization module on the basis of minimal detrimental effect to the semantic accuracy. Extensive simulation results validate that SparseSBC shows efficient and effective transmission performance under various channel conditions, and outperforms typical SemCom solutions. 
\end{abstract}

  
\section{Introduction}\label{sec1}
\IEEEPARstart{R}{ecently}, faced with the advent of the Internet of Intelligence (IoI) \cite{li_collective_2020}, semantic communications \cite{lu_rethinking_2022,lu2023semanticsempowered}, which put more emphasis on the semantic-level accuracy, emerge as a novel and promising solution and attract significant research interest, as the classical bit-level accuracy-oriented communication techniques approach the Shannon limit and fail to satisfy stringent requirements in the IoI era.
% \IEEEPARstart{I}{n} recent years, with the development of communication technology, the communication system for the use of channel has been approaching the Shannon limit. Nevertheless, the advent of the Internet of everything era also puts forward higher requirements for the ability of communication systems. In contrast to traditional communication system limited by bit-rate transmission which cannot meet the requirements of communication, system focusing on the convey of semantic becomes the most powerful competitor in future communication.
 
In line with implementation maturity to extract semantic features from the source, SemCom could be generally classified as text, speech, image and video transmission. For example, Long Short-Term Memory (LSTM) and Transformer \cite{vaswani2017attention}-based natural language processing techniques have been extensively leveraged to design a Joint Source-Channel Coding (JSCC) mechanism \cite{farsad2018deep} for text transmission \cite{xie2021deep,xie2020lite,zhou2021semantic}. Meanwhile, in order to understand emotions and tunes of speech more thoroughly, attention-based squeeze-and-excitation \cite{weng2021semantic} and symbol recognition modules \cite{shi2021new} have been added to the basic JSCC structure for speech transmission. Furthermore, as for image transmission, limited to the storage of equipment and capability of transmission channel, JSCC-based image transmission methods \cite{bourtsoulatze2019deep,kurka2021bandwidth} are often contingent on image compression works. In that regard, Convolutional Neural Network (CNN)-based \cite{prakash2017semantic,zhang2022wireless} and Recurrent Neural Network (RNN)-based methods \cite{toderici2017full} demonstrate astonishing performances in (lossy) image compression. Besides, a Generative Adversarial Network (GAN)-based method \cite{rippel2017real} achieves real-time adaptive compression. Belonging to the most challenging works, semantic video transmission takes account of the temporal relationship across frames and calibrates Deep Neural Networks (DNNs) to implement conditional coding and context learning, so as to capture essential keypoints and adapt to wireless channels \cite{jiang_wireless_2023,wang_wireless_2023}. Notably, most works adopt an end-to-end approach to train Neural Network structures, and implicitly assume the differentiability in channel layer. Nevertheless, such an assumption might not hold in practice \cite{lu_rethinking_2022}. 
% Centering on the closer distance in semantic space, Natural Language Processing (NLP)-based on Long Short-Term Memory (LSTM) and Joint Source-Channel Coding (JSCC) get off to a good start in text transmission\cite{farsad2018deep}. Meanwhile, transformer\cite{vaswani2017attention}-based\cite{xie2021deep,xie2020lite,zhou2021semantic} systems are proposed to transmit information by extracting semantic features from the source. Beyond text, speech signals contain more information like emotions and tunes, and need to be understood and extracted in a more complex way. Based on an attention mechanism employing Squeeze-and-Excitation (SE) modules, DeepSC-S\cite{weng2021semantic} improves the recovery accuracy of speech. To reduce errors in speech transmission, an additional symbol recognition module is added to the basic structure\cite{shi2021new}.

On the other hand, given the wide application of images and videos, extensive signal processing techniques are available to image and video compression. For example, JPEG and JPEG2000 utilize quantization and entropy coder to compress images for fewer-bits transmission. Meanwhile, HEVC \cite{sullivan2012overview} and MPEG adopt the hybrid coding framework based on transform coding and prediction. Furthermore, as a sub-Nyquist sampling framework, Compressive Sensing (CS) has been employed to capture the sparsity in images and boost the performance of numerous imaging applications \cite{kravets_progressive_2022}. Unfortunately, in the literature, there shed little light on explicitly investigating the sparsity into semantic image and video transmission. However, considering the successful applications of CS, it remains worthwhile to effectively combine latest sparsity-driven techniques and advanced DNNs towards SemCom on image and video transmission. 

In this paper, we put forward SparseSBC, an alternate learning based sparse SemCom framework for image and video transmission. In particular, SparseSBC involves a transmitter and a receiver to learn the transmission scheme by turns, so as to be applicable for both differentiable and non-differentiable channel. In addition, SparseSBC adapts a ``self-critic" scheme to overcome the divergence in training. Furthermore, SparseSBC deduces semantic bases which can describe all semantic embeddings with calibrated DNN-driven modules. On top of that, SparseSBC quantizes the sparsely represented bits from the space of semantic bases for transmission. Conversely, jointly taking account of sparse signal recovery and JSCC-based decoding, SparseSBC reconstructs the image from received noisy bits. 
%In this paper, based on our previous works \cite{lu2021reinforcement} for text transmission, which reaps powerful benefits of Reinforcement Learning (RL) to capably train SemCom transmitter and receiver separately, we put forward SparseSBC, an RL-based sparse SemCom framework for image and video transmission. In particular, SparseSBC involves an RL-empowered transmitter and receiver, so as to applicable for both differentiable and non-differentiable channel. Furthermore, SparseSBC deduces semantic bases with calibrated NN-driven modules. On top of that, SparseSBC quantizes the sparsely represented bits from the space of semantic bases for transmission. Conversely, jointly taking account of sparse signal recovery and JSCC-based decoding, SparseSBC reconstructs the image from received noisy bits.

In brief, the major contribution of this work can be summarized as follows:
\begin{itemize}
    \item We put forward SparseSBC, an alternate learning based sparse SemCom framework for image and video transmission, by effectively combining the concept of CS and the advance of DNNs. In particular, a ``self-critic" scheme is introduced into the training procedure to achieve better exploration and stable process. 
    \item Within SparseSBC, at the transmitter side, besides the encoder, we additionally transform semantically encoded bits into the space of semantic bases and quantize the results by building up DNN-driven non-linear mapping. Meanwhile, similar operations are applied to the receiver side as well.
    % \item We experimentally verify the existence of sparsity after mapping the semantically encoded bits into the space of semantic bases. 
    \item Through extensive simulations in various channel conditions, we validate that SparseSBC shows superior performance than JSCC and JPEG, in terms of the required transmission bits to obtain competitive reconstruction performance.
\end{itemize}

The remainder of the paper is organized as follows. In Section \ref{sec2}, we present the framework of SparseSBC, while the details of SparseSBC for image and video transmission are shown in Section \ref{sec3}. Section \ref{sec4} gives the corresponding simulation results. Section \ref{sec5} concludes the paper.
% The rest of the paper is organized as follows. In Section \ref{sec2}, we introduce the sparse semantic-base transmission system model based on typical models. The details of RL based image/video semantic transmission system are shown in Section \ref{sec3}. Section \ref{sec4} and Section \ref{sec5} are simulation results and conclusion respectively.

\section{System Model}\label{sec2}

 \begin{figure}[tbp]
    \centering
    \includegraphics[width = 0.49\textwidth]{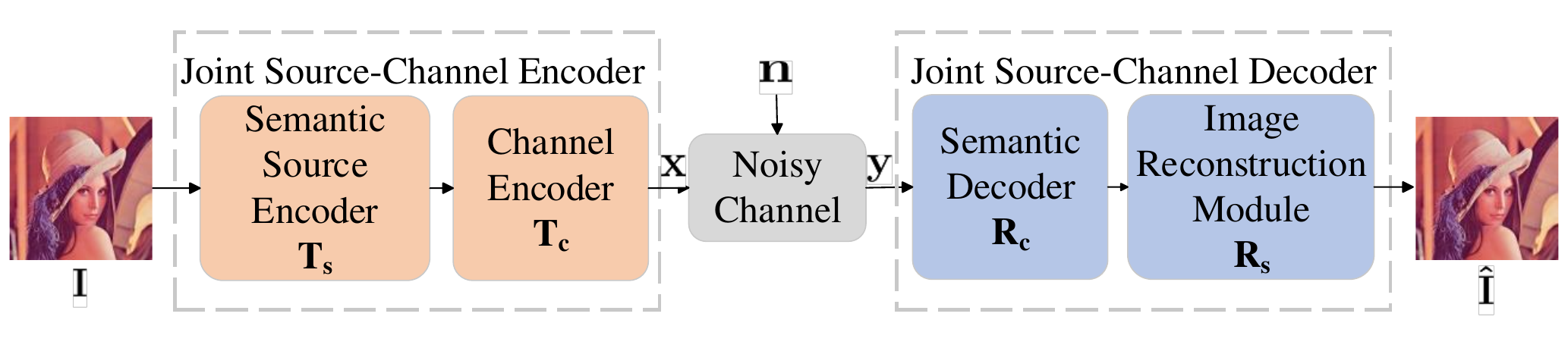}
    \vspace{-0.7cm}
    \caption{Typical block structure of SemCom. }
    \vspace{-1.8em}
    \label{fig1:a}
\end{figure}
\begin{figure*}[tbp]
    \centering
    \includegraphics[width = 1.0\textwidth]{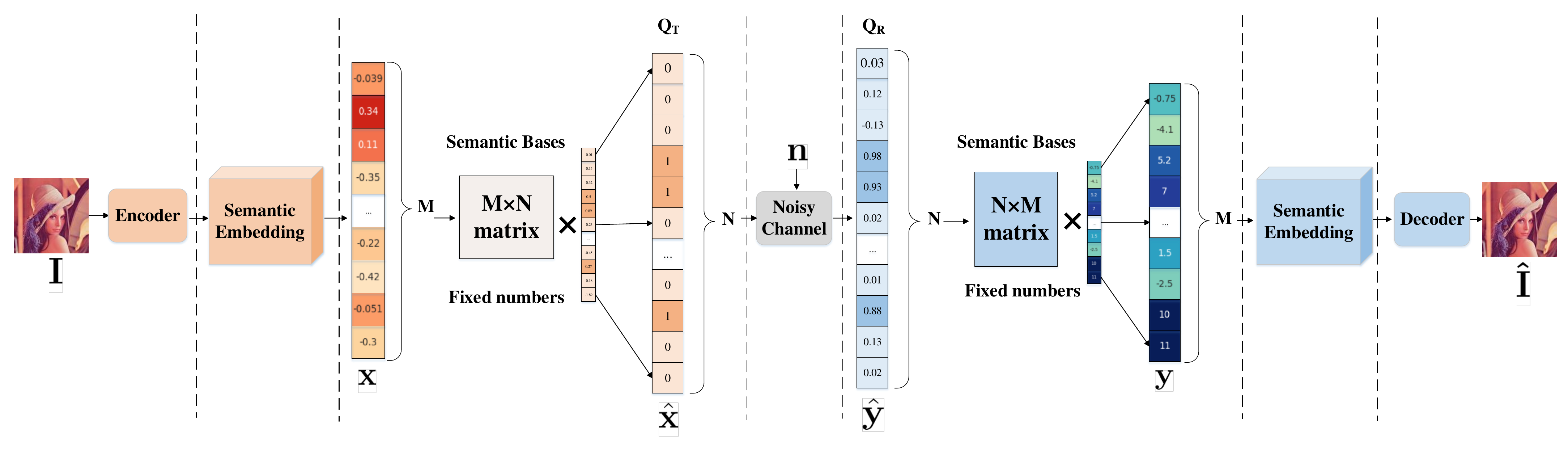}
    \vspace{-0.5cm}
    \caption{Framework of SparseSBC. }
    \vspace{-1.8em}
    \label{fig1:b}
\end{figure*}

\subsection{General Framework of SemCom}
% \subsection{Typical Semantic Communication System for Image Transmission}
% 作为JSCC的对比，从内容上感觉不应该提JSCC了
As illustrated in Fig. \ref{fig1:a}, we primarily consider the SemCom framework encompassing an encoder $\mathcal{T}$ and a decoder $\mathcal{R}$ as transmitter (TX) and receiver (RX) respectively, as well as a noisy channel $\mathcal{H}$. Both encoder and decoder are implemented by DNN and mutually contingent.
% As illustrated in Fig. \ref{fig1:a}, we primarily consider a typical semantic communication system consisting of an encoder and a decoder as transmitter (TX) and receiver (RX) respectively, as well as a noisy channel. Both encoder and decoder are mainly trained by NN and are learned jointly.
Specifically, the encoder is logically comprised of two DNN-based modules (i.e., a semantic source encoder $\mathbf{T}_\mathbf{s}$ and a channel encoder $\mathbf{T}_\mathbf{c}$) to extract low-dimensional features from the original image and encode them into symbols for channel transmission, respectively. 
% Specifically, the encoder is comprised of two NN-based modules: a semantic feature extractor $\mathbf{T}_\mathbf{a}$, which extracts low-dimensional features from the original image, and a semantic channel encoder $\mathbf{T}_\mathbf{b}$, which encodes the features into symbols for channel transmission.
Without loss of generality, an image $\mathbf{I}$, with $d_1\times d_2$ pixels, can be downscaled to lower-dimension embeddings $\mathbf{x} \in \mathbb{R}^M$ as
\begin{equation}
    \mathbf{x} = \mathbf{T}_\mathbf{c}\left(\mathbf{T}_\mathbf{s}\left(\mathbf{I}\right)\right).
    \label{eq1}
\end{equation}

Notably, video transmission can be conveniently transformed to image transmission, as a video consists of a sequence of images \cite{jiang_wireless_2023,wang_wireless_2023}. Therefore, as depicted in Fig. \ref{fig3}, after taking the first frame of image as the base background, subsequent images in a video can be obtained by further computing the difference of corresponding pixel's value between the two frames and taking the absolute values. This kind of differential images only record the moving parts in a video, and is easier to be compressed due to the inherent sparsity. Besides, in order to offset the accumulated effect of noise, we set a new frame as the latest background (i.e., the basis for differential operation) every 12 frames.

At the receiver side, the received signals $\mathbf{y}$ can be formulated as $\mathbf{y} = \mathbf{h}\mathbf{x}+\mathbf{n}$, where $\mathbf{h}$ denotes the channel coefficients, $\mathbf{n}$ represents the independent and identically distributed (i.i.d.) noise in channel, following a circularly symmetric Gaussian distribution. In that regard, the channel could be modeled as a non-trainable layer to fit the training process. Typically, the Additive White Gaussian Noise (AWGN) channel and the Phase-Invariant Fading (PIF) channel are generally considered in most typical studies \cite{lu_rethinking_2022}.
Mathematically, the channel can be considered as an AWGN channel when $\mathbf{h}$ is set to $\mathbf{1}$, while in PIF channel $\mathbf{h}$ is assumed to follow a constant distribution.

Conversely, RX utilizes a semantic channel decoder $\mathbf{R}_\mathbf{c}$ and an image reconstruction module $\mathbf{R}_\mathbf{s}$ to recover from noisy bits. Besides, with a denormalization layer, for every color channel each pixel's value is restored to $(0,255)$. Mathematically,
\begin{equation}
    \mathbf{\hat{I}} = \mathbf{R}_\mathbf{s}\left(\mathbf{R}_\mathbf{c}\left(\mathbf{y}\right)\right),
    \label{eq2}
\end{equation}
where $\mathbf{{\hat{I}}}$ denotes the reconstructed image.
Basically, to further evaluate the quality of the reconstructed image, we adopt the $L_1$-norm between pixels in the original image and reconstructed one. In other words, $\mathcal{L} = \frac{1}{2d_1d_2}\sum_{i,j}^{d_1,d_2}|\mathbf{I}_{i,j}-\mathbf{\hat{I}}_{i,j}|$, where $\mathbf{I}_{i,j}$ denotes the pixel $(i,j)$ of a $d_1\times d_2$ image $\mathbf{I}$. And the smaller $\mathcal{L}$ yields higher semantic similarity of $\mathbf{I}$ and $\mathbf{{\hat{I}}}$. 
\subsection{Framework of SparseSBC}
% \subsection{Sparse Semantic Bases in Communication System for Image Transmission}

% Information is a unity of meaning and symbols, and the inner meanings only can be expressed through external forms such as words, images and so on. Therefore, every symbol system can be regarded to a language that conveys meaning, which are in the form of semantic vectors constituting a specific semantic space.

% First and foremost, we consider about the bases, which can express all the vectors in the whole space by transformation. Extending to a semantic space $\mathcal{V}$, it seems feasible to decompose a set of semantic bases $\mathcal{B}$ to indicate different semantic vectors, which is formulated as $\mathcal{B} = \{e_1,e_2,...,e_n\}$. Thus every semantic vector $v$ can be denoted as:
% \begin{equation}
%     v = \lambda_1e_1+\lambda_2e_2+...+\lambda_ne_n,  \left(\lambda_1,\lambda_2,..\lambda_n\right)\in\mathbb{F}^n,
%     \label{eq3}
% \end{equation}
% where $\left(\lambda_1,\lambda_2,..\lambda_n\right)$ means the unique fractional representation under the bases $\mathcal{B}$. $\mathbb{F}^n$ denotes the coefficient field.

Fig. \ref{fig1:b} presents the framework of SparseSBC. Notably, compared to SemCom in Fig. \ref{fig1:a}, SparseSBC prominently adds a CS-consistent DNN module $\mathbf{Q}_T$ to further map $\mathbf{x}$ to $\hat{\mathbf{x}} \in \mathbb{Z}_2^N$, where $\mathbb{Z}_2 = \{0,1\}$. 
Since each image can be fully described by semantic bases, through the process of $\mathbf{Q}_T$, an image can be expressed sparsely in terms of a specific set of bases, and thus transformed into a sequence of bits with larger compression ratio. 
Correspondingly,
\begin{equation}
    \hat{\mathbf{x}} = \mathbf{Q}_T(\mathbf{x}) = \mathbf{Q}_T\left(\mathbf{T}_\mathbf{c}\left(\mathbf{T}_\mathbf{s}\left(\mathbf{I}\right)\right)\right).
    % \mathbf{x} = \mathbf{Q}_T\left(\mathbf{T}_\mathbf{b}\left(\mathbf{T}_\mathbf{a}\left(\mathbf{I}\right)\right)\right),
    \label{base}
\end{equation}

After receiving the sparse quantization signal $\hat{\mathbf{x}}$, RX uses another dequantization module $\mathbf{Q}_R$ to process the received signal $\hat{\mathbf{y}} = \mathbf{h}\hat{\mathbf{x}}+\mathbf{n}$ and recover the images, that is,
\begin{equation}
    \mathbf{\hat{I}} = \mathbf{R}_\mathbf{s}\left(\mathbf{R}_\mathbf{c}\left(\mathbf{Q}_R\left(\hat{\mathbf{y}}\right)\right)\right).
    % \mathbf{\hat{I}} = \mathbf{R}_\mathbf{a}\left(\mathbf{R}_\mathbf{b}\left(\mathbf{Q}_R\left(\mathbf{y}\right)\right)\right),
    \label{debase}
\end{equation}

In resemblance to CS \cite{kravets_progressive_2022}, $\mathbf{Q}_T$ and $\mathbf{Q}_R$ is somewhat equivalent to nonlinear mapping and reconstruction matrices.
Besides, for simplicity of representation, we denote 
%$\mathcal{T}=\mathbf{Q}_T\left(\mathbf{T}_b\left(\mathbf{T}_a\right)\right)$ and $\mathcal{R}=\mathbf{R}_a\left(\mathbf{R}_b\left(\mathbf{Q}_R\right)\right)$.
the transmitter as $\mathcal{T}=\mathbf{Q}_T\left(\mathbf{T}_\mathbf{c}\left(\mathbf{T}_\mathbf{s}(\cdot)\right)\right)$ and the receiver as $\mathcal{R}=\mathbf{R}_\mathbf{s}\left(\mathbf{R}_\mathbf{c}\left(\mathbf{Q}_R(\cdot)\right)\right)$ hereafter.

Apart from adopting $\mathcal{L}$ (i.e., $L_1$-norm) to measure the semantic similarity, a sparse factor $\mathcal{L}_{\text{s}}$ will be used to impose the sparsity of the transmitted bits. In other words,
\begin{equation}
    \mathcal{L}_{\text{full}} = \underbrace{\frac{1}{2N}\sum_{i,j}^{d_1,d_2}|\mathbf{I}_{ij}-\mathbf{\hat{I}}_{ij}|}_{\mathcal{L}} + \underbrace{\varepsilon \sum_{i=1}^{N}{\left|\hat{x}_i\right|}}_{\mathcal{L}_{\text{s}}},
    \label{sparseloss}
\end{equation}
where $\hat{\mathbf{x}} = [\hat{x}_1, \cdots, \hat{x}_N]$ and $\varepsilon$ denotes the sparsity weight.
% As for the objective function, we will keep the loss function as $L_1$-norm, which has a strong robustness. Meanwhile, a sparse loss function will be added into training to reduce the code weight of final bitstream for transmission: $\mathcal{L} = \varepsilon \sum_{i=1}^{N}{\left|x\right|}$, where $N$ is the number of sequence after binary quantization, $\varepsilon$ denotes the sparse weight which we set to $0.1$.

Next, we will explain how to develop modules $\mathbf{Q}_T$ and $\mathbf{Q}_R$, and use the distributed training scheme to bypass the differentiality issue of random channels.

\section{Implementation Details of SparseSBC}\label{sec3}
\subsection{DNN Structure}
In this paper, we adopt the following DNN structure and provide one viable means while leaving the utilization of other DNN structures (e.g., transformer \cite{vaswani2017attention}) as future works. Besides, we denote the parameters in the transmitter and receiver DNNs (i.e., $\mathcal{T}$ and $\mathcal{R}$) as $\phi$ and $\theta$, respectively.
\subsubsection{Design of Source \& Channel Encoder and Decoder} \

At the transmitter side, the joint encoder $\mathcal{T}$ encompasses multiple convolutional layers with LeakyReLu activation functions to extract semantic features from an image $\mathbf{I}$ and reshape to an $M$-length vector $\mathbf{x}$. Similarly, the decoder at the receiver side contains de-convolutional layers to restore the images.
\subsubsection{Design of $\mathbf{Q}_T$ and $\mathbf{Q}_R$}\

The DNN for both $\mathbf{Q}_T$ and $\mathbf{Q}_R$ consists of a fully connected layer with Tanh activation function, so as to transform semantically encoded bits to a sequence of numbers ranging from $-1$ to $1$. Afterwards, a binary quantization is performed to obtain an $N$-length binary bits vector (i.e., containing $0$s or $1$s only). Notably, such a fully connected layer resembles the mapping and reconstruction matrices (to map and reconstruct from semantic bases) in CS; while the latter Tanh function and quantization module capture non-linear transformations. 

\subsection{Alternate learning based SparseSBC}
\subsubsection{Integration of Alternate Learning and SemCom}\
\begin{figure}[tbp]
    \centering
    \subfigure[The training phase of TX. ]{\label{fig2:a}\includegraphics[width = 0.49\textwidth]{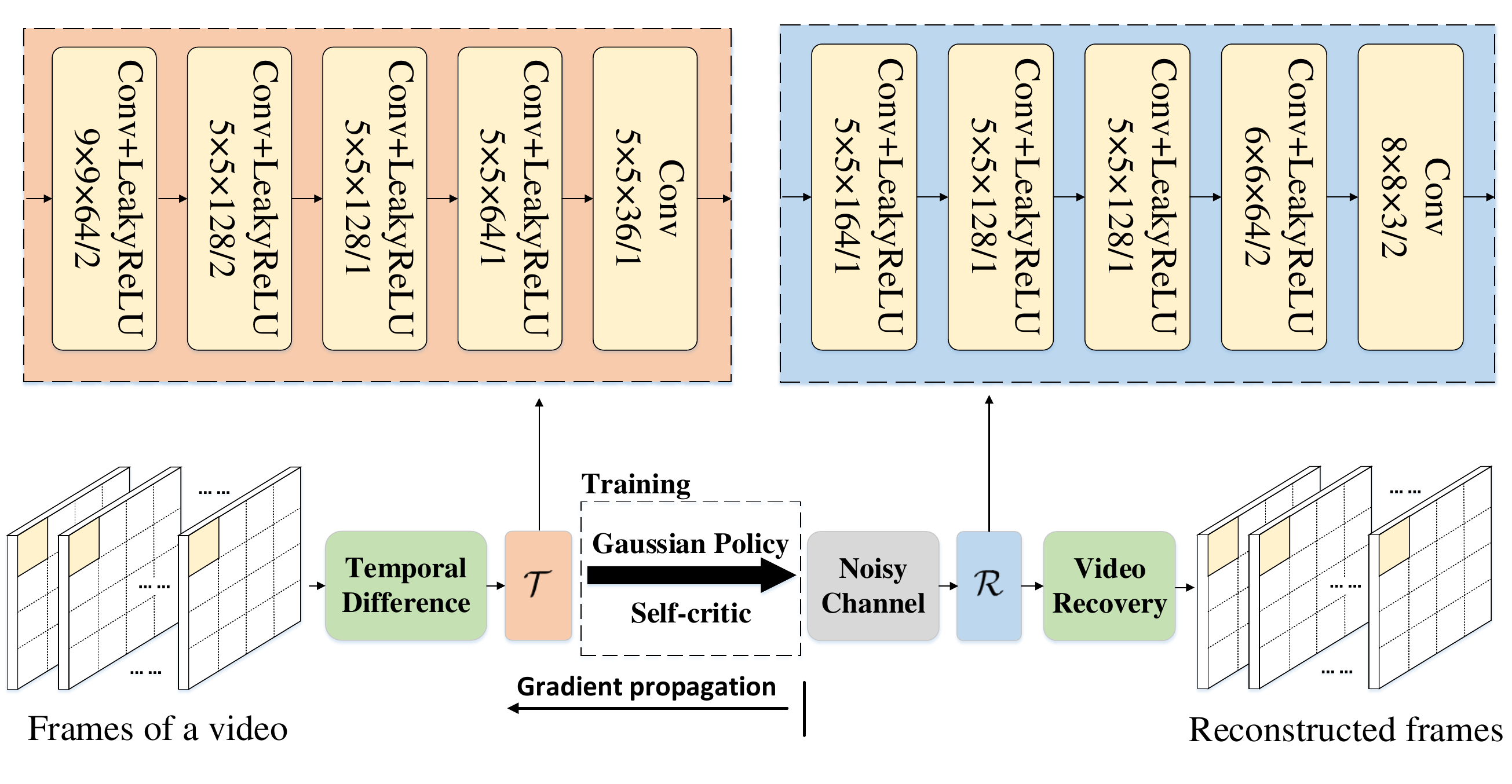}}
    \subfigure[The training phase of RX. ]{\label{fig2:b}\includegraphics[width = 0.49\textwidth]{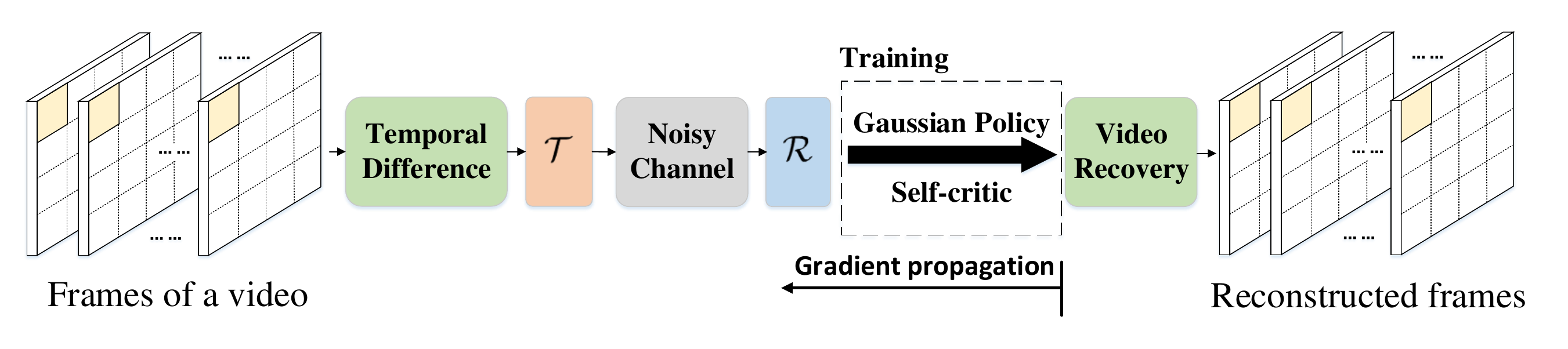}}
    \caption{The details of TX and RX in training period, which both adopt distributed training scheme. }
    \vspace{-1.8em}
\end{figure}
Most typical SemCom schemes discuss the differentiable objective optimization, but such a stringent assumption on transmission channels might not always hold in practice. Therefore, we consider an alternate learning scheme into SemCom to deal with the non-differentiability of random channels and help to turn the whole learning system into a collaborative semantic transceiver. 

To better illustrate the details of alternate learning, we define the input of $\mathcal{T}$ and $\mathcal{R}$ as $\mathbf{s}$ at time $t$ respectively. That is, for the transmitter, it encodes $s_{\text{en}}^{\left(t\right)} = \mathbf{I}^{\left(t\right)}\in \mathbb{R}^{d_1\times d_2}$ to $a_{\text{en}}^{\left(t\right)}=\hat{\mathbf{x}}^{\left(t\right)} \in \mathbb{Z}_2^N$, and at the receiver side, $s_{\text{de}}^{\left(t\right)} = \hat{\mathbf{y}}^{\left(t\right)}\in \mathbb{R}^{N}$ and $a_{\text{de}}^{\left(t\right)}=\hat{\mathbf{I}}^{\left(t\right)}\in \mathbb{R}^{d_1\times d_2}$. 
Both the transmitter and the receiver target to reconstruct images as close as the original signals. Meanwhile, the transmitter shall impose the sparsity of encoded embeddings. Hence, for the transmitter parameterized by $\phi$ and the receiver parameterized by $\theta$, the learning reward can be formulated as
\begin{subequations}
\begin{align}
    r_{\phi}^{\left(t\right)}
    &= \Theta_{\phi}(\mathbf{I},\hat{\mathbf{I}},\hat{\mathbf{x}}) 
    = 1-\mathcal{L}_{\text{full}},
    \label{eq4:a}\\
    r_{\theta}^{\left(t\right)}
    &=\Theta_{\theta}(\mathbf{I},\hat{\mathbf{I}}) 
    = 1-\mathcal{L}.
    \label{eq4:b}
\end{align}
\label{eq4}
\end{subequations}
Different from JSCC systems, in order to maximize the learning reward, we train the encoder and decoder alternately by taking a batch of $T$ samples (i.e., image transmissions) as a mini-batch to optimize the following objective function
\begin{align}\label{eq:J}
J=\mathop{\mathbb{E}}\limits_{a^{(1)},\cdots,a^{(T)}}\left[\sum\nolimits_{t=1}^{T}r^{(t)}\right],
\end{align}
where $r$ takes the formula in \eqref{eq4:b} and \eqref{eq4:a} for $\phi$ and $\theta$, respectively. 

\subsubsection{Training Procedures}\
In this part, we primarily introduce the training procedures of the transmitter in depth. As the terminology ``alternate learning'' implies, we freeze $\theta$ and thus the objective function is only parameterized on $\phi$, that is, $J\left(\phi\right)$ and $r_{\phi}$.
To further overcome the difficulty of divergence in training for high-dimensional semantic space, inspired by the methodology in Reinforcement Learning (RL), we adopt a simpler and quicker scheme named ``self-critic'' \cite{rennie2017self} based on the  Gaussian policy gradient. In particular, we attempt to learn an optimal policy $\phi^{\ast}$ for TX: 
\begin{equation}
    \phi^{\ast} = \mathop{\arg\max}\limits_{\mathcal{T}}\Theta_{\phi}(\mathbf{I},\underbrace{\mathcal{R}(\mathcal{H}}_{\text{no grad}}(\mathcal{T}(\mathbf{I})))).
    \label{eq9}
\end{equation}

Notably, instead of training the model to estimate the baseline directly, we use the average return from a group of parallel samples as the baseline \cite{rennie2017self}. Specifically, we repetitively generate the encoding results from the input $\mathbf{s}^{\left(t\right)}$ $m$ times according to the Gaussian distribution. 
% By adopting a self-critic Gaussian policy gradient, the transmitter policy can be re-formulated as
% Specifically, during the training phase of encoder $\phi$, we freeze $\theta$ and optimize $\phi$ by a self-critic Gaussian policy gradient. Therefore, the output embedding of encoder is converted to a mean value $\mu\in \mathbb{R}^{d\times1}$. The policy can be defined: 
\begin{equation}
    \hat{\mathbf{x}}^{\left(t\right)}_i \sim \operatorname{S}\left(\mathcal{N}\left(\bm{\mu}_{\phi}^{\left(t\right)}=\hat{\mathbf{x}}^{\left(t\right)},\bm{\Sigma}\right)\right), \forall i = \{1,\cdots,m\},\\
    \label{eq7}
\end{equation}
where $\operatorname{S}$ denotes the parallel sampling following the Gaussian distribution  $\mathcal{N}$. $\bm{\Sigma}=\left(\sigma^2\mathbf{E}\right) \in\mathbb{R}^{N\times N}$ is a covariance matrix set by the identity matrix $\mathbf{E}$ and a scale factor $\sigma$, which can be regarded as an exploration factor to get more abundant expression of embeddings. 
Since a constant $\sigma$ may not be well realized for exploration and exploitation, we can adjust the value of $\sigma$ in different stages of training procedure dynamically, similar with the simulated annealing algorithm, wherein the value of $\sigma$ gradually decreases along with the increase of epoch es (denoted as ``Annealed''). 
Furthermore, we also consider $\sigma=\left[\sigma_1,\cdots,\sigma_N\right]$ to be a learnable vector (denoted as ``Learnable''), which can be determined by a sigmoid function of the encoded bits, that is, $\sigma_i = \text{Sigmoid}\left(\hat{\mathbf{x}}_i\right), i\in\{1,2,\cdots,N\}$. 

Following the decoding operation at the receiver side, we can obtain $m$ rewards, that is, $\Theta_i$, $i \in \{1,\cdots, m\}$ taking a value $\Theta_{\phi}$. Without loss of generality, for any $i$, by regarding the average of the remaining $m-1$ outputs as the bias term, we calculate the difference between $\Theta_i$ and the bias term for its parameter update. On the other hand, following our previous work \cite{lu2021reinforcement}, we have the following theorem to demonstrate the result of the policy gradient propagation for TX. 
\newtheorem{theorem}{Theorem}
\begin{theorem}\label{theorem}
Let $\widetilde{\mathcal{T}}(\mathbf{I})^{\left(t\right)}$ be one of the multi-sampled embeddings in TX at time $t$. With the self-critic Gaussian policy gradient defined in \eqref{eq7}, the gradient propagation for TX is given as
\begin{equation}
\nabla_{\phi}\log\left(\pi_{\phi}^{\left(t\right)}\right) = \left[\widetilde{\mathcal{T}}(\mathbf{I})^{\left(t\right)}-\mathcal{T}(\mathbf{I})^{\left(t\right)}\right]^{\intercal} \bm{\Sigma}^{-1}\left[\nabla_{\theta}\mathcal{T}(\mathbf{I})^{\left(t\right)}\right].\label{eq10b}
\end{equation}
\end{theorem}

Therefore, the calculation of the semantic policy gradient can be summarized as
\begin{align}
    \nabla J\left(\phi\right)\approx \frac{1}{mT}\sum_{i=1}^{m}\left[\sum_{t=1}^{T}\nabla_\phi\log\pi_{i;\phi}^{\left(t\right)} \left(\Theta_i -\mathop{\text{avg}}\limits_{k\sim m;k\neq i}(\Theta_k)\right)\right],
    \label{gra}
\end{align}
where $\mathop{\text{avg}(\cdot)}$ calculates the mean value. Finally, we update $\phi$ with a learning rate $\alpha$, as 
\begin{equation}
    \phi \gets \phi - \alpha \nabla J\left(\phi\right).
    \label{update}
\end{equation}

Notably, such a ``self-critic" training scheme can alleviate the high variance problem of the plain policy gradient and keep a stable training procedure \cite{rennie2017self}. 

Similarly, in the training period of $\theta$, $\phi$ is frozen as well and the Gaussian policy gradient-based ``self-critic" is identically adopted to reconstruct images from received embeddings. Afterwards, the objective of RX can be optimized as
\begin{equation}
    \theta^{\ast} = \mathop{\arg\max}\limits_{\mathcal{R}}\Theta_{\theta}(\mathbf{I},\mathcal{R}(\underbrace{\mathcal{H}(\mathcal{T}(\mathbf{I}))}_{\text{no grad}})).
    \label{policyR}
\end{equation}

We summarize the details of SparseSBC in Algorithm \ref{alg:algorithm1}. 
 \begin{algorithm}[tbp]
  \caption{Distributed training based SparseSBC}
  \label{alg:algorithm1}
  \begin{algorithmic}[1]
  \renewcommand{\algorithmicrequire}{\textbf{Input:}}
  \renewcommand{\algorithmicensure}{\textbf{Output:}}
  \REQUIRE Batch size $T = 64$, initial learning rate $\alpha = 1e^{-4}$, self-critic samples $m = 5$, epoch $E = 200$, semantic similarity metric $\Theta$, scale factor $\sigma$\
  \ENSURE Encoder parameter $\phi$, decoder parameter $\theta$\
  \STATE{For video transmission: Repeat the  temporal-difference processing to obtain the base image (every 12 frames) and differential images}
  \FOR {$\rm{epoch}=1:E$} 
  % \STATE {Sample a batch of images}
  \STATE{\textbf{Training TX}}
    \STATE {For each batch, TX encodes each sample image into its sparse binary embedding, on the basis of frozen parameters $\theta$ at the RX. }
    \STATE{TX samples $m$ random samples according to (\ref{eq7}), and sends the encoded bitstreams through the channel.}
    \STATE{RX decodes with the semantic policy gradient (\ref{gra}), thus yielding the objective function (\ref{eq4:a}).}
    \STATE{TX takes gradient propagation towards $\phi$, and updates $\phi$ with $\alpha$ (\ref{update}).}
  \STATE{\textbf{Training RX}}
    \STATE{For each batch, TX encodes $m$ image samples into its sparse binary embeddings, based on its trained policy.}
    \STATE{TX sends encoded bitstreams through the channel.}
    \STATE{RX samples a sequence of random symbols from channel with the Gaussian distribution, decodes and calculates the objective function (\ref{eq4:b}).}
    \STATE{RX updates $\theta$ with Gaussian policy.}
  \ENDFOR
  \RETURN{the parameters $<\phi,\theta>$}
  \end{algorithmic}
 \end{algorithm}

\section{Simulation Results}\label{sec4}
In this section, we evaluate the performance of SparseSBC, and compare it with works like JSCC \cite{bourtsoulatze2019deep}, Multi-Level Semantics-aware Communication system (MLSC) \cite{zhang2022wireless} and JPEG. In particular, we adopt the popular dataset Cifar-10\cite{krizhevsky2009Learning}, which contains $60,000$ RGB images, with fixed sizes of $32\times32$, as well as a recorded video clip. 

In addition, we evaluate the performance in terms of metrics like number of transmitted bits, Peak Signal-to-Noise Ratio (PSNR) \cite{bourtsoulatze2019deep}, and Structural Similarity Index Measure (SSIM) \cite{sara2019image}. These two metrics respectively evaluate the recovered images objectively and subjectively. In particular, PSNR measures the ratio between the maximum possible power of signal and noise which corrupts the signal, and can be defined as $10 \log_{10} \frac{\text{MAX}^2}{\text{MSE}} (dB)$, where $\text{MSE} = \frac{1}{d_1 d_2}\sum_{i,j}^{d_1,d_2}\left(\mathbf{I}_{i,j}-\mathbf{\hat{I}}_{i,j}\right)^2$ denotes the mean squared-error, and $\text{MAX}$ is the maximum value of pixels in the image of interest (i.e., $255$ for 24-bit depth RGB images). On the other hand, SSIM, which is calculated as $\rho_l\big(\mathbf{I},\hat{\mathbf{I}}\big)^{\lambda_1}\cdot\rho_c\big(\mathbf{I},\hat{\mathbf{I}}\big)^{\lambda_2}\cdot\rho_s(\mathbf{I},\hat{\mathbf{I}}\big)^{\lambda_3} \in [0,1]$, captures luminance, contrast and structural differences between images by $\rho_l$, $\rho_c$ and $\rho_s$ with exponential coefficients $\lambda_1$, $\lambda_2$ and $\lambda_3$. Typical experiemental settings are summarized in Table \ref{parameter}.

\begin{table}[tbp]
    \vspace{-0.8em}
    \centering
    \caption{Simulation settings}
    \label{parameter}
    \begin{tabular}{c|c}
    \hline
    Parameter &  Value\\ 
    \hline
    Learning rate $\alpha$ & $10^{-4}$ \\
    Batch size $T$ & $64$  \\
    Sparse weight $\varepsilon$ & $0.1$ \\
    Length of $\hat{\mathbf{x}}$ $N$ & $5000$ \\
    Length of $\hat{\mathbf{y}}$ $M$ & $2304$ \\
    Self-critic samples $m$ & $5$ \\
    Scale factor $\sigma$ & $0.1$\\
    Image dataset & Cifar-10\\
    \hline
    \end{tabular}
\vspace{-1.8em}
\end{table}

\begin{table*}[tbp]
    \centering
    \caption{Comparison of SparseSBC with the traditional digital method JPEG and general SemCom system JSCC and MLSC in terms of performance and the bits of one image to be transmitted (SNR=10). }
    \label{bit}
    \begin{tabular}{c|c|c|c|c|c|c|c|c}
    \toprule
       & SparseSBC & \multicolumn{5}{c|}{General JSCC} & MLSC-image & JPEG  \\
    \midrule
    Float Resolution (Bit) & $1$ & $1$ & $4$ & $8$ & $16$ & $32$ & $32$ & $1$ \\
    \hline
    No. Transmission Bytes & $625$ & $288$ & $1,152$ & $2,304$ & $4,608$ & $9,216$ & $1,536$ & $\geq 643$ \\
    \hline
    PSNR (dB)& $27.66$ & $13.18$ & $19.40$ & $27.13$ & $27.16$ & $27.16$ & $25.44$ & $28.21$  \\
    \hline
    SSIM ($\%$) & $92.12$ & $60.79$ & $84.43$ & $97.67$ & $97.83$ & $97.83$ & $80.46$ & $94.37$  \\
    \bottomrule
    \end{tabular}
    \vspace{-1.8em}
\end{table*}

%The semantics in images can be described by an identical set of semantic bases and arbitrary real numbers in the corresponding position. Sparse semantic bases play an important role in following operations with their expressiveness and flexibility. We use PSNR to measure the quality of the recovered images by both schemes. From the results we find that sparse semantic bases reduce code weights with less performance reduction, which demonstrates advantages of sparse semantic bases. 

\begin{figure}[tbp]
    \vspace{-0.8em}
    \centering
    \subfigure[PSNR]{\includegraphics[width = 0.36\textwidth]{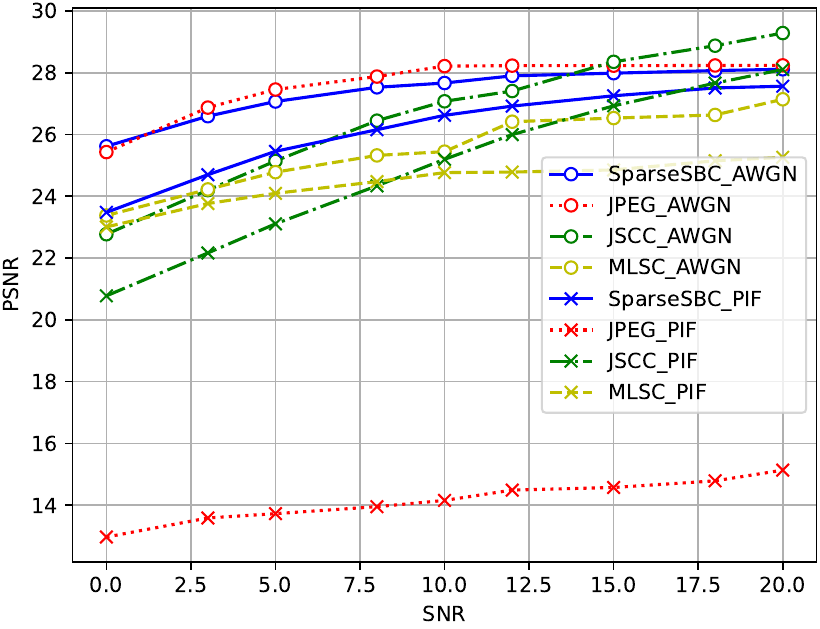}}
    \subfigure[SSIM]{\includegraphics[width = 0.36\textwidth]{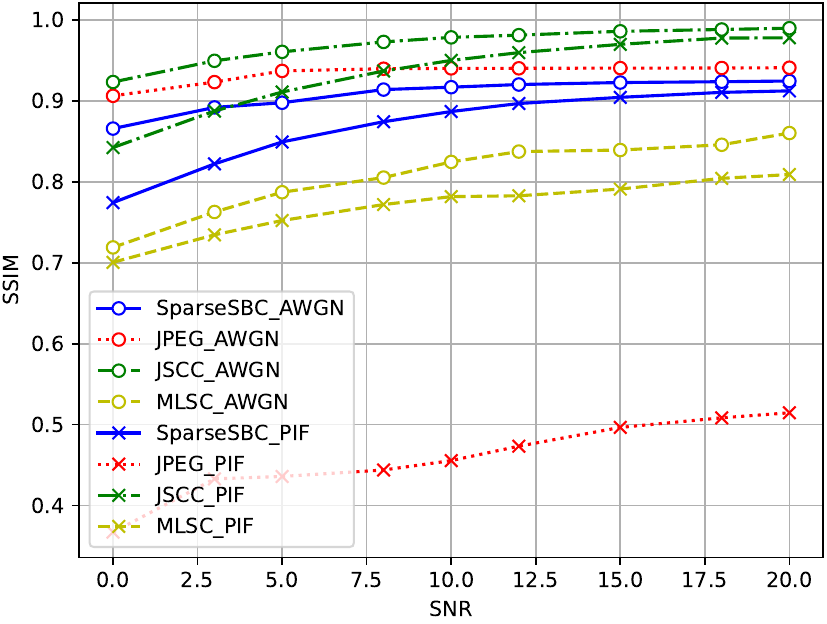}}
    \vspace{-0.8em}
    \caption{Performance comparison of SparseSBC with JPEG, JSCC and MLSC in terms of PSNR and SSIM.}
    \label{fig5}
\end{figure}

We first testify the performance of SparseSBC under different channel conditions and present the performance comparison with JPEG, JSCC and MLSC in terms of PSNR and SSIM in Fig. \ref{fig5}. 
It can be observed from Fig. \ref{fig5}, under AWGN channels, SparseSBC achieves similar performance compared to JPEG, while in PIF channels SparseSBC significantly outperforms JPEG. In other words, JPEG hardly works in harsh environments while SparseSBC retains the performance to resist poor channels. 
Meanwhile, SparseSBC yields superior PSNR than JSCC and obtains competitive SSIM as JSCC, which shows a stable performance under the influence of channel noise. Furthermore, SparseSBC leads to superior performance than MLSC, which further demonstrates its advantages. 

On the other hand, Table \ref{bit} summarizes average number of transmitted bits under different techniques. It can be observed that SparseSBC compresses every Cifar image to the fixed $625$ bytes. As a comparison, JSCC needs approximately $2,300$ bytes to approach similar PSNR as SparseSBC, while JPEG consumes more than $643$ bytes to transmit an image and requires significantly more bits along with increasing image complexity. Meanwhile, SparseSBC outperforms MLSC in both transmitted bits and performance. In a nutshell, SparseSBC is rather communication efficient. 
% Besides, we test the proposed system under different channel conditions and compare the results with the traditional separation-based digital transmission scheme JPEG, and the results are shown in Fig. \ref{fig5}. In AWGN channels SparseSBC achieves similar performance compared to JPEG, while in FIF channels our method outperforms JPEG. Therefore, SparseSBC is more robust under different channel conditions, and can always keep a steady result. When JPEG drops sharply in FIF channel, our method keeps a steady performance in both AWGN and FIF channels. 

% Meanwhile, we also count the average bits for each pixel of an image of each method. The compression ratio of JPEG for images is not a fixed value, which provides a maximum value when it comes to solid color images. Compared to more than 643 bytes for JPEG compression method of solid-color images, our method compress every Cifar image to 625 bytes. Thus, SparseSBC is proved to pay less communication cost to convey corresponding semantic information. 
\begin{figure}[tbp]
    \vspace{-1em}
    \centering
    \includegraphics[width = 0.36\textwidth]{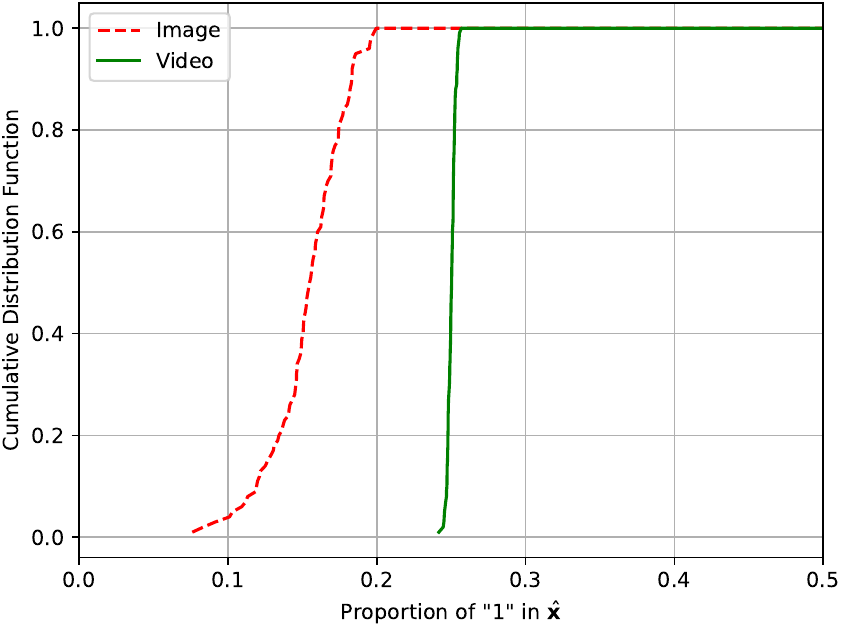}
    \vspace{-1.2em}
    \caption{The sparsity of transmitted bits in SparseSBC for image and video transmission.}
    \vspace{-1.8em}
        \label{fig6}
\end{figure}
\begin{figure}[tbp]
    \centering
    \includegraphics[width = 0.36\textwidth]{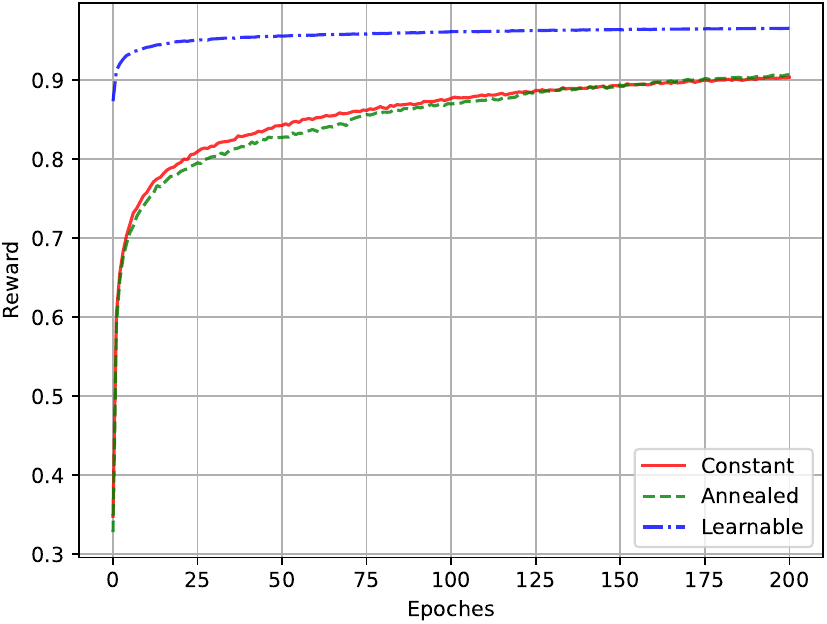}
    \vspace{-1.2em}
    \caption{The sensitivity the of scale factor $\sigma$. }
    \vspace{-1.2em}
    \label{sigma}
\end{figure}
Fig. \ref{fig6} records transmitted bits for Cifar test images, and clearly demonstrates the sparsity of transmitted bits, since ``1''s account for less than $20\%$ of whole transmitted bits. Furthermore, similar observations apply for the video transmission with a sparsity of less than $30\%$. 
We also explore the sensitivity of the scale factor $\sigma$, as shown in Fig. \ref{sigma}. It can be observed that the learnable $\sigma$ shows better performance in training period, which outperforms the constant and annealed value $\sigma$. 

\begin{figure}[tbp]
    \centering
    \subfigure[First frame]{\label{fig3:a}\includegraphics[width = 0.24\textwidth]{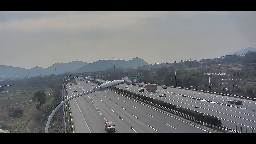}}
    \subfigure[Second frame]{\label{fig3:b}\includegraphics[width = 0.24\textwidth]{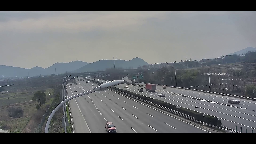}}
    \subfigure[Differential image]{\label{fig3:c}\includegraphics[width = 0.24\textwidth]{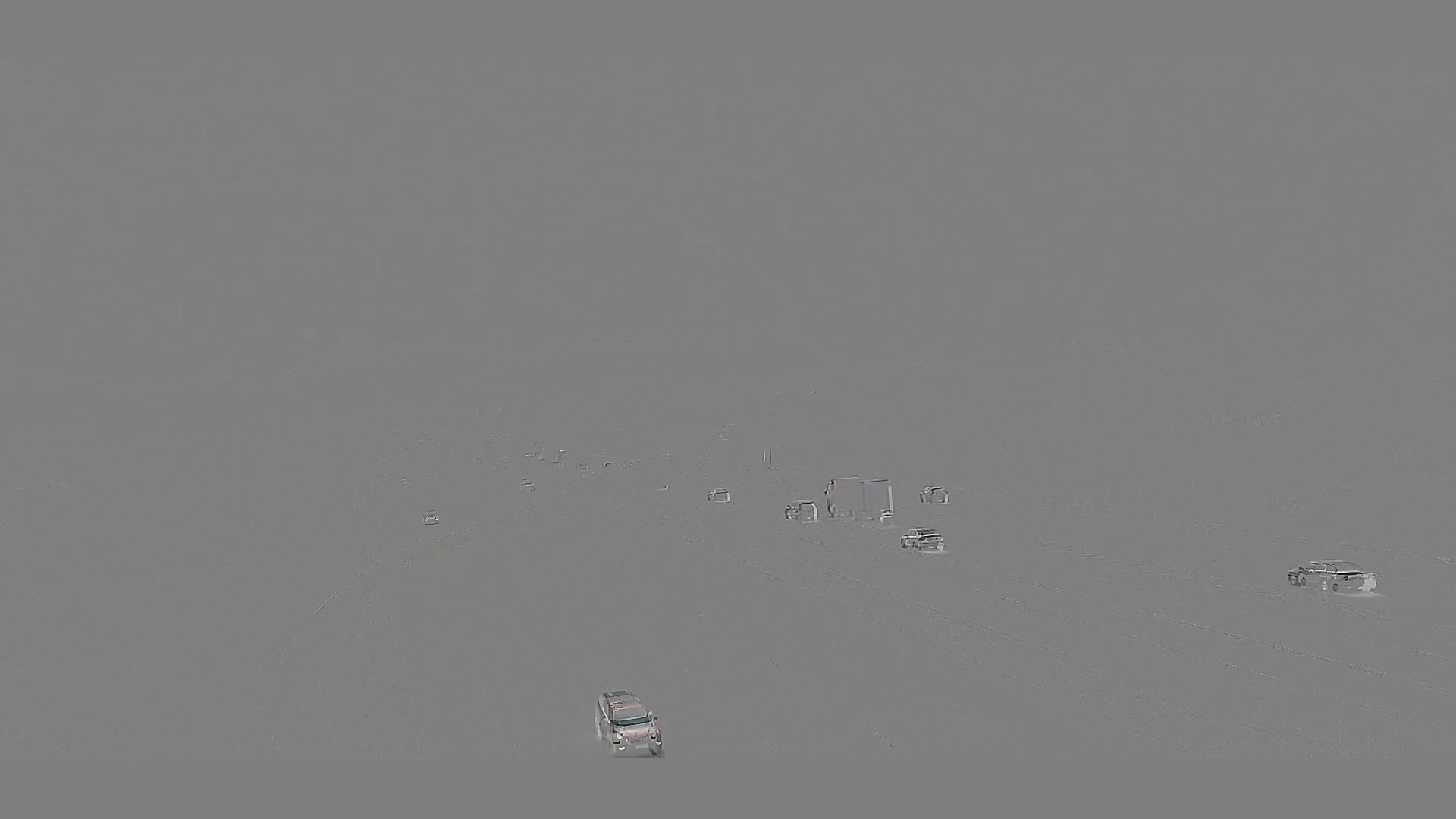}}
    \subfigure[Reconstructed second frame]{\label{fig3:d}\includegraphics[width = 0.24\textwidth]{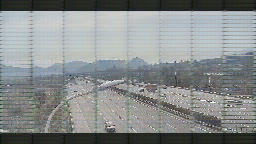}}
    \vspace{-1.2em}
    \caption{Examples of differential images extracted by temporal difference from the video clip and reconstructed video frame.\label{fig3}}
    \vspace{-1.2em}
\end{figure}
Fig. \ref{fig3} illustrates the process of temporal difference for video transmission and presents the result to reconstruct the second frame based on temporal difference-involved SparseSBC. Though quantization and channel noise lead to performance reduction, SparseSBC shows its stable performance in video transmission. 

 \section{Conclusions}\label{sec5}
In this paper, we have proposed a sparse SemCom system for visual transmission, named SparseSBC, which capably learns the DNN-based encoder and decoder deployed on the transmitter and receiver alternately, so as to adapt to the non-differentiable channel. In particular, a ``self-critic" scheme has been leveraged into the training procedure to guarantee a stable process. In addition, by extracting a set of semantic bases and implementing binary quantization, semantic information is converted into sparse bitstreams, thus effectively bridging the potential combination between semantic communications and compressed sensing. Extensive simulation results validate that for visual transmission, SparseSBC outperforms JPEG and joint transmitter-receiver transmission schemes with efficiency and effectiveness under various channel conditions. 

% \bibliographystyle{IEEEtran}
% \bibliography{reference}

\end{document}